\documentclass[a4paper]{jpconf}

\pdfoutput=1

\usepackage{graphicx}
\usepackage{color}
\begin{document}

\title{On Firewalls in quantum-corrected General Relativity}

\author{A J Nurmagambetov$^{1,}$\footnote[2]{Also at Karazin Kharkov National University, 4 Svobody Sq., Kharkov, UA 61022 \& Usikov Institute for Radiophysics and Electronics, 12 Proskura St., Kharkov, UA 61085.} and I Y Park$^3$}

\address{$^1$Akhiezer Institute for Theoretical Physics of
NSC KIPT,
1 Akademicheskaya St., Kharkov, UA 61108 Ukraine}

\address{$^3$Department of Applied Mathematics,
Philander Smith College, Little Rock, AR 72223, USA }

\ead{ajn@kipt.kharkov.ua, inyongpark05@gmail.com}

\begin{abstract}
We consider the information loss paradox at different angles, from the standard semi-classical approximation of General Relativity to the recently proposed scenarios of black holes evolution caused by effects of gravity quantization. Focusing on the Firewall proposal, we study the loop effects on the geometry and boundary conditions in black hole spacetimes and analyze the energy measured
by an infalling observer near their horizons. As a result we obtain a trans-Planckian energy transition for the time-dependent black hole solution on the quantum-induced AdS background, the importance of which for the black hole (in)formation is discussed. 
\end{abstract}

\section{Introduction}

The outstanding experimental test of the Standard Model (SM) at the Large Hadron Collider and the lack of signals of new particles and phenomena beyond the SM framework (the recent discovery of new particles by the LHCb collaboration \cite{Aaij:2018bla,Aaij:2018tnn} is another confirmation of the quark structure of mesons and baryons) gave rise to growing skepticism with respect to the so-called New Physics expected at hundreds of TeV. An increasing number of physicists is inclined to believe that the Standard Model, despite all its inherent flaws, can correctly describe the elementary particles physics even up to the Planck energies. This point of view is at least naive, because climbing up more and more closer to the Planck scale we shall observe the more and more significant contribution from the gravity force to elementary particles quantum processes. To be consistent in the whole range of energies -- from TeVs to the Planck scale -- the Standard Model has to be supplemented by General Relativity (GR), so that quantum gravity should be part of the game.
Note however, any way to involve gravitational theory within the standard paradigm of Quantum Field Theory (QFT) is faced with well-known drawbacks. Lack of unitarity is one of them.

Let us emphasise, unitarity is one of the fundamental principles of any consistent quantum theory, from Quantum Mechanics (QM) to QFT. Within the
standard approach to QM/QFT the evolution of the initial state at past infinity to the final state at future infinity, $|\Psi(-\infty)\rangle \longrightarrow
|\Psi(+\infty)\rangle$,
is completely determined by $S$-matrix:
\begin{equation}
|\Psi(+\infty)\rangle =S |\Psi(-\infty)\rangle,\qquad SS^\dag=S^\dag S=1.
\end{equation}
In particular, unitarity of $S$-matrix results in keeping the wave function probability distribution in the initial and final states,
\begin{equation}
\langle \Psi(+\infty)|\Psi(+\infty)\rangle=1=\langle \Psi(-\infty)|S^\dag S|\Psi(-\infty)\rangle\equiv \langle \Psi(-\infty)|\Psi(-\infty)\rangle, 
\end{equation}
as well as in the Principle of Detailed Balance (PDB)
\begin{equation}
 \langle \Psi(-\infty)|\Psi(+\infty)\rangle=\langle \Psi(+\infty)|S^\dag S|\Psi(-\infty)\rangle= \langle \Psi(+\infty)|\Psi(-\infty)\rangle. 
\end{equation}
The PDB implies the invariance of QM/QFT processes under $t \leftrightarrow -t$ that means any QM system always goes from the final state to the initial state upon the time returning back. Despite the probabilistic nature in part of the measurements, in part of evolution quantum mechanics obeys the Quantum Determinism: no information on the initial state is lost in temporal evolution of a quantum system, and data of measuring the final state is enough to recover the initial state.

The Standard Model, being an example of the consistent (viz. unitary) theory, drastically changes on account of gravity. One of the main results of quantum scattering theory is the so called Froissart bound \cite{Froissart:1961ux}, a consequence of which is the apparent conflict between unitarity and presence of higher-spin elementary particles: in framework of the standard QFT inclusion of non-trivially interacting elementary particles with spin higher than one inevitably results in the loss of unitarity. This drawback of the standard QFT can be overcome within the Regge pole theory, duality models of strong interactions and String Theory \cite{Veneziano:1974dr,Collins:1977,Polchinski:1998}. However, String Theory, as the main candidate for Unified Theory of matter and interactions, operates on the Planck scales. At the essentially  low energies one may ``resolve'' the unitarity problem of quantum theory of gravity and matter by use of the semi-classical approximation \cite{Birrell:1982ix}. 

Within the semi-classical approach to GR the matter fields are supposed to be quantum on a curved classical background, properties of which are determined by the classical gravitation field.\footnote[4]{The early arguments in favour of non-quantal GR are stated in \cite{Rosenfeld:1963}. The early pointed contradictions to this proposal can be found in \cite{Eppley:1977,Page:1981aj}.} Keeping this idea in mind it seems that the unitarity problem, referred to quantum gravity as a high-spin theory, will be never appeared again. However, as it was pointed by Hawking \cite{Hawking:1974sw,Hawking:1976de}, the problem is reappeared in a different way, due to the black holes (BHs) radiation flow.

At first glance it may seem that the presence of a flux of radiation from a black hole (see Fig.1) saves unitarity, which is violated due to a part of the initial quantum state fall inside the black hole. This lost portion of information could be recovered through detecting the outcoming radiation, so that the data on the final state and the collected data on the Hawking radiation will be enough to recover the initial state.

\begin{figure}[h]
\begin{center}
\includegraphics[width=11.4cm]{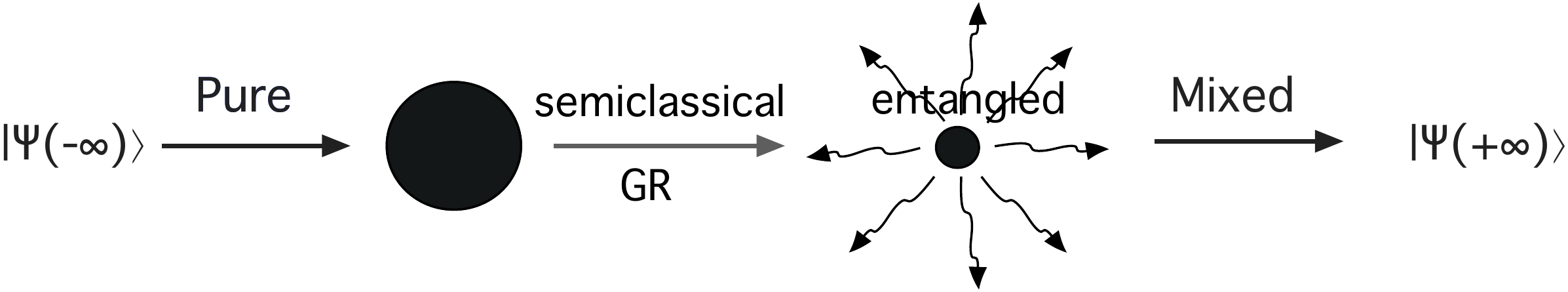}
\end{center}
\caption{\label{label}Black Hole evaporation within the semi-classical picture.}
\end{figure}

However, this rationale would be true in the case of forming the Hawking radiation by the pure states. Because the outcoming flow possesses the temperature, according to quantum statistical mechanics it has to be described by the density matrix of mixed states. Thus, the final state we will deal with will be a mixed state. It means that to restore the initial, say pure, states from the known final mixed states one needs sufficiently more information on the quantum system. For example, information on the entangled w.r.t. the Hawking radiation states inside the BH, which are completely lost in the BH evaporation process, is also required to this end. The disagreement between what one expected in the semi-classical approximation in General Relativity and what happens due to the black holes evaporation was called as the Information Paradox or, more exactly, the  ``Loss of Information'' Paradox in BH Physics. 
Further on we will briefly consider main ingredients in the Hawking picture of the BHs evolution, ways of resolving the Information Paradox and how realistic to implement one of them, the so-called ``firewall'' scenario. We end up the paper with a brief summary of the results and a concise discussion of related topics.

\section{Entanglement in QFT and GR}

As we have seen a part of the game is the entanglement of quantum states \cite{Schrodinger:1935zz}. Let us  discuss in short what the entanglement is and how it defines on quantum states.

Let us begin with consideration of two QM sub-systems that compose a whole QM system. Mixing/purity of  sub-system states strongly depends on separability of the state vector in the complete QM system. Pure/mixed  non-separable states of the complete QM system are termed as entangled states.\footnote{Under the term ``entangled states'' we silently suppose the so-called bipartite (or EPR) states.} Important to note that the projection of entangled states onto sub-systems ``$1$'' or ``$2$'' is always described by the mixed states density matrix.

To recognize the entangled states among the variety of QM states one should apply a specific criterion of 
their selection, e.g., the following one \cite{Unruh:2017uaw}: a state is the entangled state when one may establish observables ${\cal O}_{1,2}$ for the sub-systems ``$1$'' and ``$2$'', such that
\begin{equation}
\langle \Psi| {\cal O}_1 \otimes {\cal O}_2 |\Psi \rangle \ne \langle \Psi|{\cal O}_1|\Psi \rangle \langle \Psi|{\cal O}_2|\Psi \rangle .
\label{entangQFT}
\end{equation}
The physical meaning of the considered inequality (\ref{entangQFT}) is quite clear: the entanglement impacts even than parts of the common system are far remouted of each other, and any formal interaction between them is absent. Since the entanglement is one of the key characteristics of QM it also becomes crucial in its relativistic version, i.e.,  in QFT \cite{Harlow:2014yka,Unruh:2017uaw}. 

Indeed, in context of BH Physics as QFT on a curved background it becomes important the ``inverse'' effect of creating entangled states upon  partitioning a quantum system onto several sub-systems. Consider this effect on the example of the scalar QFT.

Suppose that for some physical reason (as, say, for a uniformly accelerated observer or for a gravitationally collapsing body) event horizons arise in the initially connected region of space. That means the Cauchy surface $\Sigma$ (i.e. the hypersurface of $t=\mathrm{const}$) is divided onto the pair of disjoint regions $\Sigma_{1,2}$ with the common boundary $S$: $\Sigma_1 \bigcup \Sigma_2 \bigcup S=\Sigma$ \cite{Unruh:2017uaw} (see Fig.2), cutting out of the whole space on globally hyperbolic domains $U_{1,2}$.

\bigskip
\begin{figure}[h]
\begin{center}
\includegraphics[width=7.4cm]{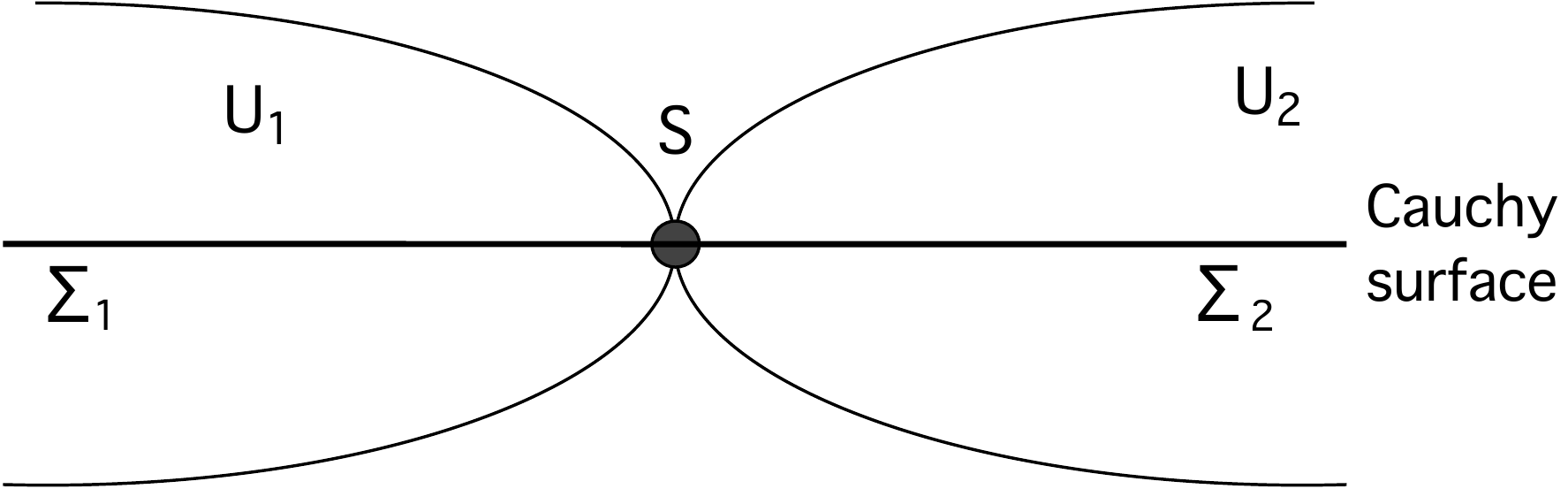}
\end{center}
\caption{\label{label}Cauchy surface splitting.}
\end{figure}
Now let's evaluate the specific criterion of the entangled states (\ref{entangQFT}) for this case.

The role of observables for the subsystems ``$1$'' and ``$2$'' is played by values ​​of the scalar field operator itself at various spatial points of the domains $U_{1,2}$. Then, the l.h.s. of (\ref{entangQFT}) is just the correlator (2-point function) over an admissible state, which is \cite{DeWitt:1965}
\begin{equation}
\langle\Psi | \phi(x_1) \phi(x_2)|\Psi \rangle \sim \frac{U(x_1,x_2)}{\sigma(x_1,x_2)}\,,\quad x_1 \in \Sigma_1,\,\,\, x_2 \in \Sigma_2 .
\end{equation}
Here $U(x_1,x_2)$ is a smooth function of the arguments; $\sigma(x_1,x_2)$ is the so-called geodesic interval ($1/2$ of the square of the distance between points $x_1$ and $x_2$ along the geodesic).

Since according to its definition
\begin{equation}
\lim_{x_{1,2} \rightarrow x} \sigma(x_1,x_2)=0, \,\,\,x \in S \,,
\end{equation}
the 2-point function diverges in the coincidence limit:
\begin{equation}
\lim_{x_{1,2} \rightarrow x} \langle\Psi | \phi(x_1) \phi(x_2)|\Psi \rangle \longrightarrow \infty .
\end{equation}
On the other hand, for any admissible state, square of the mean value of the field operator, being the right hand side of (\ref{entangQFT}) in the coincidence limit, is restricted from above:
\begin{equation}
\langle\Psi|\phi(x_1)|\Psi\rangle \langle\Psi|\phi(x_2)|\Psi \rangle \stackrel{x_{1,2} \rightarrow x}{\longrightarrow} \left(\langle \Psi|\phi(x)|\Psi \rangle \right)^2 < \infty .
\end{equation}
Therefore, the inequality (\ref{entangQFT}) holds.

To sum up, we observed the effect of entanglement of states after dividing the original system onto sub-systems. Since considering this effect we did not use any special properties of a specific QFT and evaluations were performed by use of well defined quantities in any field theory, 
this effect will take place in any physically reasonable QFT. 

One may notice that the Fig.2 looks alike as the diagram of flat space-time in the Rindler coordinates.
This similarity is not an accident, since a remarkable example of the QFT states entanglement is the Unruh effect \cite{Fulling:1972md,Davies:1974th,Unruh:1976db}. 
The appearance of mixed states with the Unruh non-trivial temperature 
\[
T_U=\frac{\hbar a}{2\pi c k_B} \,,
\]
which are detected by a uniformly accelerated observer after splitting the Minkowski space onto four Rindler wedges can be explained as the manifestation of the QFT states entanglement in causally related domains \cite{Iso:2017qbo,Higuchi:2017gcd}.\footnote{Note that, like many effects in gravitational physics, the Unruh effect is a coordinate-dependent effect. This circumstance gave reasons to doubt that such an effect exists (see the original paper \cite{Narozhny:2001PRD} and the subsequent comments on that \cite{Fulling:2004xy}, \cite{Narozhny:2004PRD}). Some of the points of criticism of \cite{Narozhny:2001PRD,Narozhny:2004PRD} were ovecome in a recent enlightening paper \cite{Louko:2018pij}. The final point in the discussion of the (non)existence of the Unruh effect can only be put experimentally (see, e.g., a recent proposal of \cite{Cozzella:2018byy}).}

The Rindler space is a toy model to describe the space-time geometry near the BH event horizon. Hence, it is inevitable for any QFT the occurrence of standard entanglement of quantum states inside and outside of BH, the consequence of which is the Hawking radiation of mixed/thermal quantum states.  (See more on the entanglement in BHs from different points of view  in \cite{Eisert:2008ur}, \cite{Braunstein:2009my} and \cite{Hutchinson:2013kka}.)

Folding up all together, the BHs evolution within the semi-classical approximation comes as follows \cite{Unruh:2017uaw} (cf. Fig.2 therein). Black holes evaporate due to the Hawking radiation, however quantum states on a late-time Cauchy surface 
after the BH evaporation are still entangled with the early-time quantum states inside the BH, 
even if they are completely disappeared during the evaporation process. That is why the initial pure quantum state turns into the final mixed state, according to that the complete knowledge on the final state turns out to be insufficient for determining the initial state. In this sense we talk on the {\it information loss} in BHs. This conclusion holds true even in more sophisticated scenarios (such as ``Many-worlds/Multiverse'', ``Baby Universe creation at the singularity point'' etc.), as long as we do not go beyond the semi-classical approximation.

\section{Firewall as a way of resolving the information loss paradox}

Armed with understanding the basic mechanisms behind the Information Paradox, 
we may consider different ways of their overcoming. Being inherently related to the semi-classical approximation, the Information Paradox most likely may not be resolved in framework of a non-radical (in terminology of \cite{Unruh:2017uaw}) proposal. Therefore, we have to consider ``radical'' alternatives that go beyond the semi-classical approximation. The most popular among them are: unforming the black hole at all due to the collapsing matter explosion before emergence of any horizon \cite{Mersini-Houghton:2014zka}; unforming the black hole at all due to converting the collapsed matter energy into ``fuzzballs'' \cite{Mathur:2005zp,Skenderis:2008qn}; forming a black star instead of the black hole \cite{Kawai:2017txu}. And finally, the remaining proposal we will focus on below is a placed at the BH horizon ``firewall'' \cite{Almheiri:2012rt} (and naturally associated with ``firewalls'' concept of energetic curtains \cite{Braunstein:2009my}).

 As we have already seen, a source of the problem is the entanglement of quantum states across the BH horizon. Disentangling the quantum states could be an important step towards the settlement of the problem. The proposal of \cite{Almheiri:2012rt} employs this observation and consists in the following. Perhaps, the Black Hole, formed in the expected way, further behaves radically different from the standard semi-classical picture by Hawking. Specifically, at some point in time at the location of the event horizon it appears an object  -- a ``firewall'' -- whose role is to ``disentangle'' the quantum states outside and inside of the BH. The offered AMPS scenario in particular employs the ``no-cloning'' theorem of quantum physics and the entropy sub-additivity of complex systems. Then, by use of these tools, one may observe that in addition to the indicated in \cite{Eppley:1977,Page:1981aj} inconsistencies of semi-classical GR there is another basic theoretical contradiction in between main postulates of the BH Physics.

Indeed, the AMPS analysis \cite{Almheiri:2012rt} of four complementarity postulates of 
\cite{Susskind:1993if}, based on the description of the BHs evolution within the {\it standard QFT} and {\it semi-classical GR}, results in inconsistency of two of them with the postulate on unobserving something extraordinary by a freely falling observer under crossing the event horizon. Therefore,  to keep the standard description of the BHs evolution within the semi-classical GR it is required, at least for the ``old'' enough BHs (with the age more than the Page time \cite{Harlow:2014yka,Page:1993}), the appearance of an extraordinary structure at the horizon. The BH horizon becomes the ``firewall'' location point, all objects falling inside the BH are burned in the firewall and any information on them disappears. As a result,  the quantum states outside and inside of the BH are disentangled, so the final quantum state after the BH complete evaporation is a pure state. 

As any other ways of resolving the information paradox, the firewall proposal contains weak points. The main objects of criticism are \cite{Unruh:2017uaw}:
\begin{itemize}
\item
The proposal by AMPS is a radical one that determines by a sharp enough deviation from the semi-classical description even in the small curvatures regime. In particular, to minimize the entanglement effect between the inside and outside of BH states it requires a singularity of quantum fields  on the horizon, which is converted into the ``firewall''.
\item
The appearance of a local space-time peculiarity in some restricted space-time domain contradicts to the spirit of GR, where all points of the space-time continuum are supposed to be equivalent. 
The formation of a ``firewall'' is comparable to the effect of an unexpected ma\-te\-ri\-a\-li\-za\-tion of the wall on the way of any uniformly accelerated observer.
\item
In addition, the process of the ``firewall'' formation within the AMPS proposal manifestly violates causality, since to localize this object exactly on the event horizon, it is necessary to know in advance the entire subsequent evolution of the system.
\end{itemize}

We are not aimed here at setting up a more fundamental basis to the Firewalls which will be free of the aforementioned handicaps. Rather, we are interested in the fundamental possibility of implementing this scenario, to the discussion of which we now turn.

\section{Realisation of Firewalls in models of black holes}

Since the Firewall is intended to ``disentangle'' quantum states inside and outside of the BH, it is guaranteed to happen with ultra-high energies/temperatures on the horizon. On account of the Tolman relation for the local temperature,
\begin{equation}
T_{\mathrm{local}}=|g_{00}|^{-1/2} T_H ,
\end{equation}
such an effect is principally achievable.

However, the more exact characteristic is the local energy density measured by a freely-falling observer
\cite{Lowe:2013}
\begin{equation}
\rho \equiv \bar{T}_{\mu\nu} U^\mu_K U^\nu_K,
\label{rhodef}
\end{equation}
where $U^\mu_K$ is the observer 4-velocity. $\bar{T}_{\mu\nu}\equiv \langle K|\hat{T}_{\mu\nu}|K\rangle$ is the v.e.v. of the energy-momentum operator over the Kruskal state $|K\rangle$ (the Hartle-Hawking vacuum \cite{Hartle:1976tp,Candelas:1980zt}) and the subscript $K$ is related to the Kruskal coordinates. As usual, the classical stress-energy tensor contains contributions of matter fields and of the cosmological constant
\begin{equation}
T_{\mu\nu}=-\frac{2}{\kappa^2}\Lambda g_{\mu\nu}+T_{\mu\nu}^{\mathrm{ matter}},
\label{Tmnclass}
\end{equation}
and it becomes a quantum operator $\hat{T}_{\mu\nu}$, entering the r.h.s. of the Einstein equations, 
\begin{equation}
R_{\mu\nu}-\frac{1}{2} g_{\mu\nu}R=\langle \mathrm{vac}|\hat{T}_{\mu\nu}|\mathrm{vac}\rangle ,
\label{semiclassEins}
\end{equation}
once we turn to the semi-classical description of gravity with matter. Within the semi-classical approximation the local energy density (\ref{rhodef}) does not undergo of strong changes on the even horizons of static charged BHs \cite{Park:2017dib}.

A short inspection of both sides of eq. (\ref{semiclassEins}) reveals an internal inconsistency of the semi-classical Einstein equations: the l.h.s. of (\ref{semiclassEins}) is proportional to the inverse of a characteristic length $l^2$, associated with the Einstein tensor, whereas the r.h.s. of the same relation contains the inverse of the Plank length square $l^2_{Pl}=\hbar G/c^3$. To consider both sides on equal footing one should either consider GR in the strong curvatures regime or to turn to the mean value of the Einstein tensor operator after the gravity quantization. In practice, both of these requirements mean going beyond the classical description of GR.

One may
go beyond the semi-classical approximation with taking into account quantum (perturbative) corrections to the Einstein GR. For (one-)loop corrections we get
\[
T_{\mu\nu}=-\frac{2}{\kappa^2}\Lambda g_{\mu\nu}+T_{\mu\nu}^{\mathrm{ matter}}+g_{\mu\nu}\left[c_1 R^2-(4c_1+c_2)\nabla^2 R+c_2 R_{\rho\sigma}R^{\rho\sigma}\right]
\]
\begin{equation}
-2\left[2c_1 R R_{\mu\nu}-(2c_1+c_2)\nabla_\mu \nabla_\nu R-2c_2 R_{\rho\mu\nu\sigma}R^{\rho\sigma}+c_2 \nabla^2 R_{\mu\nu}+\dots \right],
\label{semiquantT}
\end{equation}
where, following \cite{Boulware:1985wk}, we have included the loop corrections into the effective stress-energy tensor.\footnote{Note that corrections to the classical stress-energy tensor (\ref{Tmnclass}) of the second-order in curvatures or derivatives (cf. eq. (\ref{semiquantT})) are induced by the backreaction of the quantized matter and quantum graviton fields.}

For the quantum-induced description of GR (at the one-loop corrections level) the local energy density at the horizons of static charged BHs in flat spacetime -- Melvin-Schwarzschild (MS) solution (BH in an external magnetic field) and generalized MS solution \cite{Preston:2006ze} -- does not change much. A jump of the local energy density near the horizon can be realized for:
\begin{itemize}
\item
a non-stationary BH solution with matter fields;
\item
space-times with non-trivial (negative) cosmological constant induced by quantum effects;
\item
special boundary conditions, specified in the way that the quantum matter fields modes take non-trivial values on the AdS boundary (the non-Dirichlet b.c.).
\end{itemize}
Once all these conditions are fulfilled \cite{Nurmagambetov:2018het},  the local energy density undergo a trans-Planckian jump 
\begin{equation}
\rho \sim 1/\kappa^2=1/16\pi G\sim E^2_{\mathrm{Pl}}.
\end{equation}

\section{Summary and open questions}

To summarize, we have discussed the inherent relation of the Information Paradox to the semi-classical approximation in gravitational Physics: the loss of unitarity in classical gravity interacting with quantized matter is a consequence of shortcomings of the approach. At the same time, we should note that all the negative aspects of the semiclassical approximation are fully compensated by its main positive hallmark: a gravitational system evolves predictively, its dynamics is governed by second order differential equations. Nevertheless, the status of the Hawking paradox as a long-standing problem and the absence, despite of the numerous attempts, of its solution within the semi-classical approximation encourages searches for resolving the problem beyond the framework of the chosen approach. A natural way to this end is to take into account
the loop perturbative corrections to the standard gravitational action, coming either from the direct quantization of GR \cite{Park:2018vci} or from String Theory and its effective low-energy supergravity actions \cite{Polchinski:1998}. 

Going beyond the semi-classical approximation we have shown a principle possibility of arising the quantum-induced trans-Plankian energy densi\-ti\-es near the BHs horizons \cite{Nurmagambetov:2018het} at their birth and early stages of evolution. It means that we have found something different to the original AMPS proposal, since the Firewall prescription of \cite{Almheiri:2012rt} supposes the formation of this object at the late stages of the BH evolution. However, we should note that the coincidence with the AMPS Firewall can be reached for the Super-Massive Black Holes (SMBHs) in the Galactic nuclei, where the accretion disks of the surrounding matter are large enough to support the Firewalls up to the Page times. Another observation in favour of the quantum-induced trans-Planckian energy density at the SMBHs event horizons is the detected ultra-hard radiation from the active Galactic nuclei, that could also be explained within the Firewall concept. All of these observations deserve close attention and additional studies. (See \cite{Park:2019lkh,AJNYIP:2019} for further steps on this way.)

The same concerns to open questions naturally related to our research. For example, the observed by us jump of the local energy density on the horizon may be treated as a phase transition. So the question is: what is the order of the phase transition, if any? Another question is about the role of the quantum induced cosmological constant. The recent results with exact nonstationary BH solutions in Minkowski space \cite{Booth:2018xvb,Baccetti:2018otf} are against the Firewalls (there were not observed traces of trans-Planckian transitions). It would be interesting to understand the impact of analyticity of the highly idealized Vaidya solution on this result, and the possibility to realize Firewalls for non-analytic BH solutions in flat and dS spaces. We hope to address these and other open questions in our future studies.

\end{document}